\documentclass{article}
\usepackage{graphicx} 
\usepackage[a4paper, total={6in, 9.5in}]{geometry}
\usepackage{amsmath,amsthm,amsfonts}
\usepackage{float}
\usepackage{bbm}
\usepackage{booktabs}
\usepackage{siunitx}
\usepackage{url}
\usepackage{todonotes}
\usepackage{changes}
\usepackage[hidelinks]{hyperref}
\title{Gaussian mixture models for model improvement}

\author{Paolo Villani$^{1,2}$ \and Daniel Andres Arcones$^3$
        \and Jörg F. Unger$^3$ \and Martin Weiser$^2$}
\date{March 2026}

\newtheorem{definition}{Definition}
\newtheorem{remark}{Remark}
\newcommand{\R}{\mathbb{R}}
\newcommand{\N}{\mathbb{N}}

\newcommand{\dirichlet}{\mathrm{Dir}}

\newcommand{\param}{{\boldsymbol{\theta}}}

\begin{document}

\maketitle

\footnotetext[1]{Freie Universität Berlin, Arnimallee 6, 14195 Berlin}
\footnotetext[2]{Zuse Institute Berlin, Takustr. 7, 14195 Berlin}
\footnotetext[3]{Bundesanstalt für Materialforschung und -prüfung, Unter den Eichen 87, 12205 Berlin}

\begin{abstract}
Modeling complex physical systems such as they arise in civil engineering applications requires finding a trade-off between physical fidelity and practicality. Consequently, deviations of simulation from measurements are ubiquitous even after model calibration due to the model discrepancy, which may result from deliberate modeling decisions, ignorance, or lack of knowledge.

If the mismatch between simulation and measurements are deemed unacceptable, the model has to be improved. Targeted model improvement is challenging due to a non-local impact of model discrepancies on measurements and the dependence on sensor configurations. Many approaches to model improvement, such as Bayesian calibration with additive mismatch terms, gray-box models, symbolic regression, or stochastic model updating, often lack interpretability, generalizability, physical consistency, or practical applicability.
    
This paper introduces a non-intrusive approach to model discrepancy analysis using mixture models. Instead of directly modifying the model structure, the method maps sensor readings to clusters of physically meaningful parameters, automatically assigning sensor readings to parameter vector clusters. This mapping can reveal systematic discrepancies and model biases, guiding targeted, physics-based refinements by the modeler. The approach is formulated within a Bayesian framework, enabling the identification of parameter clusters and their assignments via the Expectation-Maximization (EM) algorithm.

The methodology is demonstrated through numerical experiments, including an illustrative example and a real-world case study of heat transfer in a concrete bridge.

\vspace{1ex}
\noindent\textbf{Keywords:} model deficiency, model improvement, mixture models \\
\textbf{MSC 2020:} 62F15, 62H30, 62P30, 93B30
\end{abstract}

\section{Introduction and state of the art}\label{sec:sota}
Modeling complex physical systems is inherently challenging because the appropriate level of detail, both on the level of the geometry and on the level of the physical phenomena to be considered, depends on the specific problem setup and the quantity of interest. Most models are parameterized, and these parameters are closely tied to the simplifications in the model. For example, representing a reinforced concrete bridge as an isotropic material effectively homogenizes the properties of concrete and reinforcement to an average value that depends on the actual geometry and is therefore unknown. Consequently, identifying suitable model parameters based on experimental data is a critical step in developing a reliable model~\cite{KaipioSomersalo2005,Tarantola1987}.

Once the parameters have been identified, the next step is to compare simulation results with experimental measurements to validate the model. If the model is found to be deficient, i.e. not to reproduce the observed measurements within their expected measurement error range, one can either focus on uncertainty quantification or on model improvement. Uncertainty quantification accepts the deficient model but aims at providing a good estimate of the achieved accuracy. If the model is to be used for making predictions, it is of vital importance to also quantify the prediction errors. This is, e.g., the goal of embedding methods~\cite{andresarconesBiasIdentificationApproaches2024,andresarconesModelBiasIdentification2024,sargsyanEmbeddedModelError2019} and stochastic model updating~\cite{biRoleBhattacharyyaDistance2019,maresStochasticModelUpdating2006}. Here, a stochastic model is built with an output distribution that matches the distribution of observations sufficiently well. This is achieved by replacing deterministic model parameters by random variables and identifying their distributions. Such methods are most appropriate for situations where multiple, nominally identical but randomly varying systems are to be described by a single model, but are also used for uncertainty quantification in the case of model form uncertainty. These methods are non-intrusive, but limited to situations where the measurements lie close to the output manifold of the deterministic forward model.

In contrast, model improvement approaches aim at refining the model such that it can reproduce the measurements better. This is, however, far from straightforward. Discrepancies between simulated predictions and measurements are often not directly linked to errors at the same local position in the model; they may originate elsewhere, depending on the governing equations and sensor configuration. Determining why a discrepancy occurs and which part of the model requires refinement is usually an iterative process based on trial and error and expert judgment.

Approaches to treat model form uncertainty and to improve simulation models based on some observations or measurements are manifold. Common to all is that they rely on an enrichment of the simulation model in one way or the other. 

In a seminal paper, Kennedy and O'Hagan~\cite{kennedyBayesianCalibrationComputer2001} propose to include an additive model response mismatch term with a Gaussian process prior, and to identify it in a Bayesian setting along with the original model parameters. Plumlee~\cite{plumleeBayesianCalibrationInexact2017} proposes a variant where the mismatch term is orthogonal to the model response surface, arguing that as much as possible of the observations should be explained by the original model and not by the mismatch term.
Essentially, this amounts to a sequential identification of the model parameters and the mismatch term. This is a straightforward and non-intrusive method, which, however, neither provides an improved physical model understanding nor allows predicting model states or responses not directly related to the available observations. A more fine-grained control over the mismatch term, at the expense of a more intrusive approach, is obtained by gray box models, where mismatch terms are added to individual constitutive equations in a PDE model~\cite{bruderBlackboxesBayesianInverse2018}. Still, the interpretability of the results and their generalization capabilities strongly depend on the level of detail of the model mismatch terms. 

Methods aiming directly at extending the model by interpretable and physically consistent terms include symbolic regression~\cite{Langley1977BACONAP,schmidt2009distilling} and the more recent sparsity-constrained dictionary search~\cite{bruntonDiscoveringGoverningEquations2016,CARDERERA2025116675,chen1998atomic,stephanyPDELEARNUsingDeep2024}. These approaches rely on finding the simplest (by number of additional terms in the model equations) extension by selecting the most suitable terms from a potentially huge dictionary. While providing excellent interpretability and true discovery potential, they suffer from the need to define the dictionary in the first place, from its combinatorically growing size, and from the need to have a solver at hand that is able to cope with all these extensions, or arbitrary subsets, at once. Consequently, this kind of approach is highly intrusive. Less intrusive are attempts to learn from data just particular aspects of the model, such as constitutive laws~\cite{beckPerspectiveMachineLearning2021,fuhgReviewDataDrivenConstitutive2025,maLearningNeuralConstitutive2023}, with the risk of missing the real cause of model deficiency.

Mixture models haven been used widely, in particular in social sciences and economics, to identify models where the unobserved association to a certain subgroup of the sample determines the statistical properties of the observation~\cite{lubkeInvestigatingPopulationHeterogeneity2005,xiaoMixtureDistributionAnalysis1999}, including outlier detection~\cite{aitkinMixtureModelsOutliers1980} and image smoothing~\cite{polzehlAdaptiveWeightsSmoothing2000}. The focus is generally on the identification of the latent subgroups as cofounding factors, and selecting an appropriate number of subgroups through model comparison~\cite{grimmModelFitComparison2021}. Related, but less firmly grounded in stochastic models, are mixture of experts approaches~\cite{jacobsAdaptiveMixturesLocal1991} which aim to combine multiple models to construct a better overall model. 

Many of the previously discussed approaches for model improvement or error detection attempt to enlarge the model by introducing additional terms, sometimes automatically. Although these methods increase flexibility and improve data fitting, they often lack generalizability, are cumbersome to apply, and rarely provide interpretable insights that guide physics-based model refinement.

In this work, we propose a different approach. Rather than directly modifying the model structure, we aim to understand model discrepancies by mapping complex, often hard-to-interpret sensor readings onto clusters of physically meaningful model parameters using a mixture model approach. This mapping, including the automatic assignment of sensor readings to parameter clusters, is learned within our procedure, eliminating the need for manual intervention in the analysis. The resulting structure, which determines which sensors correspond to which parameter cluster and provides the associated parameter values, offers interpretable insights into systematic discrepancies and model bias. These insights indicate where and why the model could be improved, enabling users to make targeted, physics-based refinements.

\paragraph{Notation.} By $\mathbbm{1}$ we denote the vector with all entries equal to 1. The Kronecker-$\delta$ is written as $\delta_{i,j}$. The $k$-dimensional probability simplex $\{\omega\in\R^{k+1} \mid \|\omega\|_1 = 1 \}$ is denoted by $\Delta_k$.

\section{Problem definition}
Let us first define the parameter identification problem formally and introduce the related terminology.

\begin{definition}
    The \emph{physical reality} is (for our purposes) a function  
    \begin{equation}
        f: \Xi \to \R
    \end{equation}
    mapping an \emph{input} $\xi \in \Xi$ to a real quantity $f(\xi)$ that can be measured. The input space $\Xi$ is usually a product space of \emph{sensor readings} associated to loads, time or experiment number, environmental and operational conditions, as well as sensor index or position.
    
    A computational \emph{model} of $f$ is a parametrized mapping 
    \begin{equation}
        \tilde f: \Xi \times \Theta \to \R
    \end{equation}
    intended to reproduce measurements if the parameter $\param\in\Theta\subset\R^m$ is chosen optimally, i.e. $\tilde f(\cdot,\param_*) \approx f(\cdot)$.
\end{definition}

We call a computational model deficient if it cannot explain all sensor readings in the range of their error statistics \emph{with a single parameter $\param\in\Theta$}. We define the notion of "explaining all sensor readings" in terms of the two-sided Kolmogorov-Smirnov (KS) tests~\cite{lehmannTestingStatisticalHypotheses2008,raoLinearStatisticalInference1973}, with the null hypothesis that the calibrated model $\tilde f$ agrees completely with the real system $f$. For simplicity of exposition, we restrict the consideration to independent normally distributed measurement errors, but point out that any error statistics can be treated in the same way.

Let $\xi \in \Xi^n$ be a sequence of inputs and $y\in\R^n$ a family of corresponding sensor readings $y_i = f(\xi_i) + \epsilon_i$ subject to normally distributed measurement errors $\epsilon_i\sim\mathcal{N}(0,\sigma_i^2)$ with mean zero and known standard deviation $\sigma_i$. For a given parameter vector $\param$ and the normalized residuals $\sigma_i^{-1}(y_i - \tilde f(\xi_i,\param))$, we consider the empirical discrete cumulative distribution function (CDF) of their squares:
\begin{equation}
    \hat F_{\param}(\eta,\sigma) := \frac{1}{n} \left| \big\{ i\in[1,n]\cap \N \mid (y_i-\tilde f(\xi_i,\param))^2 < \eta\sigma_i^2\big\}\right|.
\end{equation}
This is just the fraction of sensor readings for which the normalized magnitude of the residual is less than $\sqrt{\eta}$. If the model $\tilde f(\cdot,\param)$ agrees with the reality $f$, the CDF $\hat F_\param$ should coincide with the CDF $F_{\chi^2_1}$ of the normalized measurement errors, which in the chosen setting of normally distributed errors $\epsilon_i$ is the $\chi^2$ distribution for one degree of freedom. The KS test is based on the statistic
\begin{equation} \label{eq:KS-statistic}
    D(\param) := \sup_{\eta\in\R_+} |F_{\chi^2_1}(\eta)-\hat F_\param(\eta,\sigma)|, 
\end{equation}
which is asymptotically distributed as the Kolmogorov distribution $K$, i.e. $\sqrt{n}D \sim K$ for $n\to\infty$. The test rejects the null hypothesis of a model statistically indistinguishable from reality with a significance level of $\alpha\in\mathopen]0,1\mathclose[$, i.e. an upper bound on the probability of false negatives, if 
\begin{equation}
    F_K(\sqrt{n} D(\param)) > 1-\alpha, \quad \text{where} \quad
    F_K(\lambda) := \begin{cases} 0, & \lambda \le 0 \\
                                  1-2\sum_{k=1}^\infty (-1)^{k-1} e^{-2k^2\lambda^2}, & \lambda >0 \end{cases}
\end{equation}
is the CDF of the Kolmogorov distribution $K$.

\begin{remark}\label{rm:one-sided-KS}
    In cases where only an upper bounds $\bar\sigma_i\ge\sigma_i$ on the noise's variances is known, instead of the symmetric KS statistic~\eqref{eq:KS-statistic} one can consider the one-sided KS-distance 
    \begin{equation}
    \bar D^-(\param,\bar\sigma) = \sup_{\eta\in\R_+} F_{\chi^2_1}(\eta)-\bar F_\param(\eta,\bar\sigma),
    \end{equation}
    which only counts too large residuals.     
    The one-sided KS test then rejects the null hypothesis of a correct model at significance level $\alpha\in\mathopen]0,1\mathclose[$, if 
    \begin{equation}\label{eq:model-deficiency-B}
       \inf_{\param\in\Theta}  F^-_K\left(\sqrt{n}  \bar D^-(\param,\bar\sigma) \right)  > 1 - \alpha
    \end{equation}
    holds, where $F_K^-(\lambda) = 1-e^{-2\lambda^2}$ is the one-sided KS limit.  
    The one-sided test only detects deficiency due to underfitting, and does not report overfitting as an issue -- in agreement with the fact that the actual noise variance $\sigma_i^2$ might be much smaller than its known upper bound $\bar\sigma_i^2$.
\end{remark}

In practice, it makes little sense to expect the model $\tilde f$ to be statistically indistinguishable from the reality, even if calibrated. We therefore accept a certain deviation of the CDFs of size $d$, defining first an extension of the KS which allows for a deviation level $d$.

\begin{definition}[model deficiency]\label{def:deficiency}
    A model $\tilde f$ is called \emph{$(\delta,\alpha)$-deficient with deviation level $\delta \in\mathopen]0,1\mathclose[$ and significance level $\alpha\in\mathopen]0,1\mathclose[$}, if 
    \begin{equation}\label{eq:model-deficiency-A}
       \inf_{\param\in\Theta}  F_K\left(\sqrt{n}  (D(\param)-\delta) \right)  > 1-\alpha
    \end{equation}
    holds. 
\end{definition}

One drawback of model deficiency as of Def.~\ref{def:deficiency} is that it is a test with binary result and does not immediately allow to tell which of two deficient models is better. 
Moreover, the deviation level $\delta$ in the test might be challenging to fix a priori. Thus, we introduce the model deviation $d$ of the model as a quantitative measure for discrepancy, utilizing the extended KS test given by~\eqref{eq:model-deficiency-A}.
\begin{definition}\label{def:deviation}
    With the notation of Def.~\ref{def:deficiency}, the model deviation is the quantity
    \begin{equation}\label{eq:model-deviation}
        d = \inf\left\{ \delta \in\mathopen]0,1\mathclose[ \, \Big| \, \inf_{\param\in\Theta}  F_K\left(\sqrt{n}  (D(\param)-\delta) \right)  \leq 0.95 \right\},
    \end{equation}
    i.e. the minimum deviation level so that the model is deemed non-discrepant with significance level $\alpha = 0.05.$
\end{definition}

\begin{remark}
    The model deficiency condition~\eqref{eq:model-deficiency-A} and the model deviation quantity~\eqref{eq:model-deviation} are equivalent to exact global optimization over $\Theta$. In practice, one will resort to local gradient-based optimization for finding a maximum likelihood estimate or maximum posterior estimate $\param_*$, and call the model deficient if~\eqref{eq:model-deficiency-A} holds, computing the corresponding deviation level. Multistart can help in avoiding bad local minimizers. Alternatively, Markov Chain Monte Carlo sampling can be used to cover $\Theta$ thoroughly.
\end{remark}

For models with a great discrepancy from the data, the model deviation $d$ itself is also not a particularly intuitive quantity to compare models: if the residuals due to model deficiency are large compared to the measurement error standard deviations  $\sigma_i$, then $d\approx 1$. As an idealized example, we consider a model with residuals distributed according to $\epsilon_i \sim \mathcal{N}(0,a^2\sigma_i^2)$, i.e. the model deficiency is that the residuals are by a factor $a>1$ larger than expected. In that case, for significant amounts of data $d$ behaves as 
\begin{equation} \label{eq:asymptotic-KS-distance}
d \approx 1-\frac{2}{1+\sqrt{a}},
\end{equation}
with a relative error of less than 4\%. Therefore, we define the inverse function of~\eqref{eq:asymptotic-KS-distance}, i.e. $a(d)$, as a quantitative measure of large model deficiency.
\begin{definition}
    With the notation of Def.~\ref{def:deviation}, let 
    \begin{equation}
        \mathcal{D}= \left( \frac{1+d}{1-d} \right)^2 -1
    \end{equation}
    be the \emph{model discrepancy}. 
\end{definition}

Note that if the model is deemed non-discrepant at significance level $0.05$, i.e. $\tilde f$ satisfies
\[
 \inf_{\param\in\Theta}  F_K\left(\sqrt{n}  D(\param) \right)  \leq 0.95
\]
and passes the KS test, then both the model deviation $d$ and the model discrepancy $\mathcal D$ are zero. $\mathcal{D}\approx 0$ is a good agreement, while $\mathcal{D}\gg 0$ signals a mismatch larger than expected, i.e. a deficient model. \newline \medskip

A deficient model cannot reproduce the measured data with a single parameter vector $\param$. Making the model more flexible by using more than one parameter vectors for different subsets of experiments may allow to fit the data better. Then, the assignment of sensor readings to parameter vectors, the dependence of the parameter vectors on inputs, and the reduction in data mismatch can provide hints to possible causes for model deficiency. In the following, we formulate the model extension in terms of mixture models and apply it to both artificial and real model improvement problems.

\section{Mixture models for discrepancy analysis}

For analyzing model deficiency, we consider mixture models as a non-intrusive extension of the original model $\tilde f$, and as a way to generate some insight into possible causes for model deficiency.

\subsection{Bayesian inversion for mixture models}\label{sec:BI4GMM}

\begin{definition}
    A $k$-cluster \emph{mixture model extension} of a computational model $\tilde f$ for a set $\Xi^n$ of inputs is a parameterized mapping
    \begin{equation}\label{eq:mixture-model}
        \hat f: \Xi^n \times \{1,\dots,k\}^n \times \Theta^k \to \R^n, \quad 
        (\xi,z,\underline \param) \mapsto \left( \tilde f(\xi_i,\param_{z_i})\right)_{i=1,\dots,n}
    \end{equation}
    that for each sensor reading $i$ selects one of the $k$ parameter vectors contained in $\underline \param = (\param_1, \dots, \param_k)$. The set of sensor readings assigned to parameter vector $\param_j$, i.e.\ $\{ i \mid z_i = j\}$, is called the \emph{cluster} $j$.
\end{definition}
Given inputs $\xi\in\Xi^n$ and corresponding noisy sensor readings $y\in\R^n$, we formulate a Bayesian inverse problem for the parameter vectors $\underline \param\in\Theta^k$ and the assignment $z$ of inputs to parameter vectors.
We assume as in Def.~\ref{def:deficiency} independent Gaussian noise, such that $y_i=f(\xi_i)+\epsilon_i$ with $\epsilon_i \sim \mathcal N(0,\sigma_i^2)$. The likelihood is therefore 
\begin{equation}
    L_y(z,\underline \param) := \prod_{i=1}^n \exp\left(-\frac{1}{2\sigma_i^2}(y_i-\tilde f(\xi_i,\param_{z_i}))^2\right) \propto \pi(y\mid z,\underline\param)  .
\end{equation}
The prior is constructed from a given prior $\pi(\param)$ for the parameter vectors $\param_j\in\Theta$ and a categorical product distribution $P(z,\omega)$ for the assignments $z$:
\begin{equation}
    \pi(z,\underline \param\mid \omega) :=  \prod_{i=1}^n \omega_{z_i} \prod_{j=1}^k \pi(\param_j) .
\end{equation}
Here, $\omega\in\Delta_{k-1}$ provides the prior probabilities $\omega_j$ of sensor readings being assigned to one of the parameter vectors $\param_j$, and is therefore closely related to the cluster size.
As this is a priori unknown, we treat the probabilities $\omega_j$ as hyper-parameters to be identified along with $z$ and $\underline \param$, and impose a Dirichlet hyper-prior 
\begin{equation} \label{eq:hyper-prior}
    \pi(\omega) = \dirichlet(a,\omega) \propto \prod_{j=1}^k \omega_j^{a_j-1}
\end{equation}
with $a\in\R_+^k$. Since the parameter vectors $\param_i$ are a priori completely interchangeable, we choose a symmetric hyperprior with $a=\gamma\mathbbm{1}$
for some $\gamma>0$. 

Note that $\gamma$ expresses our belief in how equal the clusters are in size. A large value of $\gamma$ leads to similar cluster sizes, whereas small values allow clusters to have rather different sizes. Choosing $\gamma=1$ yields a uniform prior on $\omega$.

The posterior probability density is, by Bayes' theorem, proportional to the product of likelihood, prior, and hyperprior:
\begin{align} \label{eq:full-posterior}
    \pi(z,\underline \param,\omega \mid y) 
    &\propto \pi(y\mid z,\underline \param)\,\pi(z,\underline \param\mid \omega)\, \pi(\omega) \notag \\ 
    &\propto \prod_{i=1}^n \omega_{z_i} \exp\left(-\frac{1}{2\sigma_i^2}(y_i-\tilde f(\xi_i,\param_{z_i}))^2\right)
             \, \prod_{j=1}^k \pi(\param_j) \omega_j^{\alpha-1}.
\end{align}
We will also be interested in marginalizing the posterior~\eqref{eq:full-posterior} over $z$,
\begin{align}
    \pi(\underline \param,\omega \mid y) 
    &\propto \sum_{z\in\{1,\dots,k\}^n} \prod_{i=1}^n \omega_{z_i} \exp\left(-\frac{1}{2\sigma_i^2}(y_i-\tilde f(\xi_i,\param_{z_i}))^2\right)      \,\prod_{j=1}^k \pi(\param_j) \omega_j^{\alpha-1} \notag\\
    &= \prod_{i=1}^n \sum_{j=1}^k \omega_{j} \exp\left(-\frac{1}{2\sigma_i^2}(y_i-\tilde f(\xi_i,\param_{j}))^2\right)       \, \prod_{j=1}^k \pi(\param_j) \omega_j^{\alpha-1}, \label{eq:marginal-posterior}
\end{align}
and in the conditional posterior distribution for $z$,
\begin{equation}
    \pi(z\mid \underline \param,\omega,y) \propto \prod_{i=1}^n \omega_{z_i} \exp\left(-\frac{1}{2\sigma_i^2}(y_i-\tilde f(\xi_i,\param_{z_i}))^2\right).
\end{equation}

\subsection{Fitting mixture models}

In order to obtain both the parameter values $\underline \param$ and the assignment $z$ of sensor readings to parameters in an easily interpretable way, we aim at computing point estimates. We may either compute the maximum posterior estimate (MAP) of the full posterior~\eqref{eq:full-posterior}, or the MAP of the marginal posterior~\eqref{eq:marginal-posterior} for $\underline \param$ and $\omega$ along with a conditional posterior of the assignments $z$.

A maximizer of the marginal posterior~\eqref{eq:marginal-posterior} can be computed by the expectation maximization (EM) algorithm~\cite{dempsterEM}. This classic iterative algorithm fits a simpler lower bound model, the evidence lower bound (ELBO) to the marginal posterior such that they agree in the current iterate $(\param^l,\omega^l)$, and maximizes that lower bound for obtaining $(\param^{l+1},\omega^{l+1})$, consequently converging to a (local) maximizer of the marginal posterior. The MAP of the full posterior~\eqref{eq:full-posterior} can similarly be computed by the classification EM algorithm~\cite{celeuxClassificationEMAlgorithm1992}, alternating between $z$ and $(\underline \param,\omega)$. The MAP computation in $z$ leads to a hard clustering of experiments. Different clusters are associated with different parameters $\param_j$. In contrast, the EM algorithm provides the conditional posterior distribution $\pi(z\mid \underline \param,\omega,y)$ in form of a product categorical distribution, which can be interpreted as a soft clustering with the probability 
\begin{equation}\label{eq:z-posterior}
P(z_i = j) = \frac{\omega_{j} \exp\left(-\frac{1}{2\sigma_i^2}(y_i-\tilde f(\xi_i,\param_{j}))^2\right)}{\sum_{l=1}^k \omega_{l} \exp\left(-\frac{1}{2\sigma_i^2}(y_i-\tilde f(\xi_i,\param_{l}))^2\right)}
\end{equation}
giving the degree to which sensor reading $i$ belongs to cluster $j$.

The MAP estimates are not unique, as at least permutations of the clusters provide a multitude of -- equivalent -- global maximizers. In addition, further local maximizers are often encountered, in which both the standard EM algorithm and the clustering EM algorithm can get stuck. Multi-start techniques may be necessary to find a global or at least a sufficiently good local maximizer.

\subsection{Interpreting mixture model fits}
\label{sec:interpretation}

The insights for causes of model deficiency and corresponding options for model improvement provided by the mixture model fitting depend, of course, highly on the particular model and type of problem. Here, we focus on parameter identification problems for partial differential equation models and discuss possible causes of model deficiency and expected corresponding outcomes of fitting a mixture model to the actual measurement data.

\textbf{Changing environmental conditions.} Environmental conditions may change over time or with the batch of test specimens. This should ideally be modeled as input $\xi_i$ varying between different sensor readings, where the reading number $i$ encodes, among other aspects, also time or the test specimen used. This will lead to a clustering of sensor readings according to time or batch, with modified parameters $\param_i$ making up for the different conditions. Alternating patterns hint at periodic changes such as a day-night rhythm or seasonal changes. Block patterns with rather monotone changes in the parameter may be caused by slowly varying environmental conditions or by  degradation processes such as fracture damage in brittle material, wear in contact problems, or plasticity. We will encounter an example of this type in Sec.~\ref{sec:bridge}. In contrast, random patterns without apparent structure are likely caused by fluctuations in environmental conditions that are either random or are small-scale effects that have so far not been measured or taken into account.

\textbf{Parameter dependencies and nonlinearities.} Unanticipated state-dependent model behavior will cause parameter changes corresponding to changes in loads as provided in the inputs or to changes in the system state itself. From the assignment of sensor readings to clusters one may infer whether this is a biphasic behavior as encountered in tensile tests with elasto-plastic material, a hysteresis loop, a smooth nonlinear response as in hyperelastic materials, or temperature dependence of material parameters. We will investigate an example of this type in Sec.~\ref{sec:illustrative}.

\textbf{Sensor reading failure or bias.} Often, several measurements are taken at the same time, with different sensors encoded in the readings number $i$. Failure of a single sensor, miscalibration, wrongly specified sensor position, or highly local effects such as material heterogeneity, will likely lead to all readings from this sensor to end up in a separate cluster, with a large difference in the estimated parameter value. We will encounter such local effects in Sec.~\ref{sec:bridge}.

\textbf{Sources.} Unanticipated sources may drive the system to regions in the state space that simply cannot be reached by changing model parameters. An example would be a violation of the maximum principle for diffusion problems where only the diffusivity is to be estimated. In contrast to the above reasons for model deficiency, the introduction of more parameters by a mixture model will barely reduce the residual. At the same time, the optimizer can drive the identified parameters of the mixture model to unrealistic values in a futile attempt to reduce the data mismatch. Such results therefore hint at an unmodeled effect that can shift the system state off its parameterizeable range.

The possible reasons for model deficiency and their impact on the outcome of fitting a mixture model are, of course, not unambiguously related. Nevertheless, they are correlated to each other, and therefore the mixture model fitting results can provide helpful hints for the modeler as to how and where to improve the model.

\subsection{Number of clusters}

The number $k$ of clusters in the mixture model~\eqref{eq:mixture-model} usually has a considerable impact on the computational effort, on the resulting accuracy or deviation level of the resulting model, and on the interpretability of the results. We consider two strategies for selecting an appropriate value for $k$.

First, if the mixture model approach is employed only for generating insight into possible sources of model deficiency as discussed in Sec.~\ref{sec:interpretation}, a small to moderate number of clusters will usually be sufficient. We suggest starting with $k=2$ and, if no plausible source of deficiency, such as parameter dependence on environmental conditions or model nonlinearities, are apparent, increasing $k$ slowly until a structure in the clusters' assignment and associated parameter values can be detected. 

If, in contrast, the mixture model shall serve as an improved model itself, $k$ must be chosen differently, balancing goodness of fit and model complexity. Besides standard model selection criteria like the Akaike criterion or Bayesian information criterion, the model deficiency itself can be used for selecting $k$. With known measurement error distribution we may rely on the deficiency notion in Def.~\ref{def:deficiency} also detecting overfitting as model deficiency, and find $k$ by minimizing the deviation level $\delta$ for a given significance level $\alpha$. Otherwise, one might prescribe an acceptable deviation level $\delta^*$ and find the smallest $k$ such that the mixture model is not deficient under the one-sided KS test mentioned in Remark~\ref{rm:one-sided-KS}.

Note that using a mixture model directly as improved model has certain limitations: First, it does not provide physical insight into non-modeled processes and effects, and will likely be restricted in its extrapolation power, and second, it is restricted to the used set of measurements or sensors, preventing the use of more data or predicting future outcomes. We therefore focus on the use of mixture models for hinting at causes for model deficiency.

\section{Numerical experiments}\label{sec:exp}

In this section, we will investigate the use of mixture model fitting for obtaining clues about possible reasons for model deficiency. We consider both a simple illustrative model as well as a 2D finite element model, and situations where the model deficiency is known by construction and real situations where it is unknown.

The EM algorithm has been implemented in Python, performing a fixed number of iterations and using the implementation of the Nelder-Mead method from \texttt{scipy.minimize} to minimize the negative log marginal posterior~\eqref{eq:marginal-posterior} in every M-step; the parameter space is rescaled to the unit square and the Nelder-Mead algorithm is set to perform a maximum of 50 iterations, stopping if the difference between parameter values at consecutive steps has a norm less than $\texttt{atolx} = 3 \cdot 10^{-3}$.
The code as well as data for the examples is available on Zenodo.\footnote{\url{https://doi.org/10.5281/zenodo.18432463}}

\subsection{Illustrative example} \label{sec:illustrative}
The first example is a simple analytical model for illustrative purposes.
The ground-truth reality is assumed to be 
\begin{equation}
    f:[0,1] \rightarrow \R, \quad
    \xi \mapsto \xi + \frac{1}{3} \sin{\pi \xi}, 
\end{equation}
which could be the strain for given stress in a nonlinear uniaxial tension test with hyperelastic material locking, or the stress for given strain in an elasto-plastic uniaxial tension test.

We consider the affine computational model
\begin{equation}
    \tilde f : [0,1] \times \Theta \rightarrow \R, \quad 
    (\xi, a,b ) \mapsto a \xi + b
\end{equation}
with parameter space $\Theta =  \R_+^2$.
A number $n=40$ of measurement locations $\xi_i$ is selected randomly from a uniform distribution over $[0,1]$, and the corresponding sensor readings $y_i$ are generated by adding independent noise  $\mathcal N(0,\sigma^2)$ with $\sigma = 0.05$ to the physical reality $f(\xi_i)$.
A prior for $a$ and $b$ is assumed on $\mathbb R^2_+$, consisting of two independent truncated standard Gaussians $\mathcal N (0,1).$

As shown in Tab.~\ref{tab:discrepancy-ex1}, the computational model $\tilde f$ is unable to represent the data. We consider a 2-clusters mixture model extension with a hyper-prior for the assignment probability as discussed in Sec.~\ref{sec:BI4GMM} with $\gamma = 1$, i.e., a Dirichlet distribution $\text{Dir}(\mathbbm{1}, \omega)$.
We initialize the EM algorithm with randomly generated values drawn from the prior and perform 50 iterations. Within each iteration, the optimization in the M-step is treated as described above. 

\begin{table}
    \centering
    \begin{tabular}{lcc}
        \toprule
        Model  & Deviation $\mathcal D$ &  KS statistic $D$ \\
        \midrule
         Linear regressor & 1.36 & \num{0.43} \\
         2-clusters mixture model & 0 & \num{0.01}  \\
         4-clusters mixture model & 0 & \num{0.01} \\         
         \bottomrule
    \end{tabular}
    \caption{Model discrepancy for the illustrative example. 
    For known variance, we adopt the definition of model discrepancy based on the two-sided Kolmogorov-Smirnov test introduced in Def.~\ref{def:deficiency}.
    The parameter $\param_*$ where the discrepancy is evaluated is the MAP estimate. 
    }
    \label{tab:discrepancy-ex1}
\end{table}

\begin{figure}
    \centering
    \includegraphics[width=\linewidth]{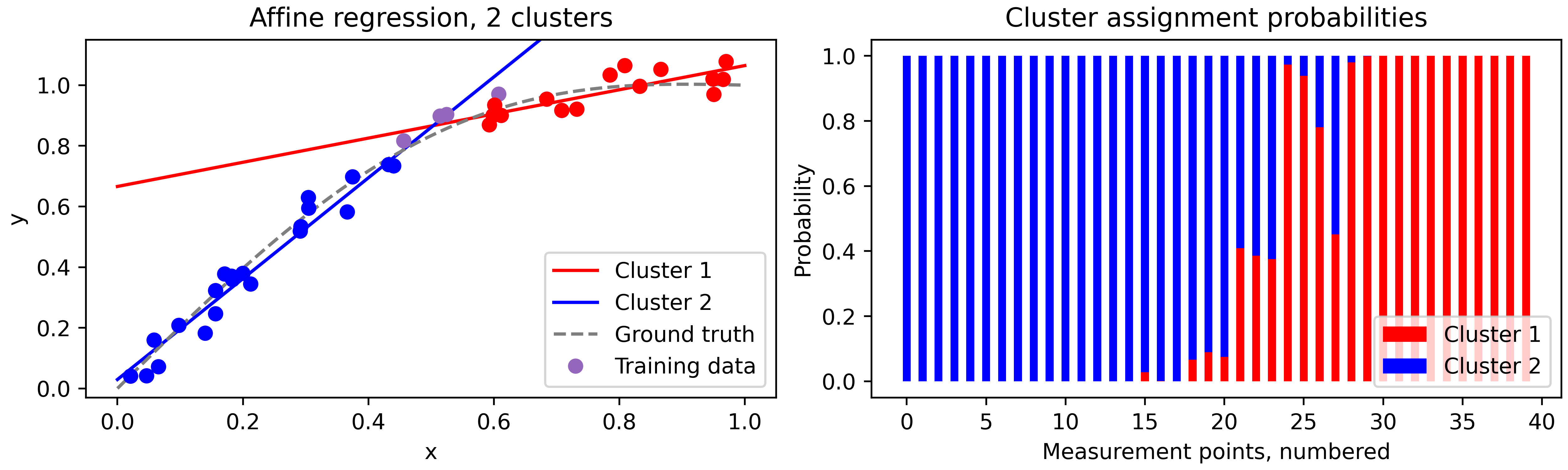}
    \caption{Results after the EM algorithm terminated for the illustrative example with 2 clusters. 
    On the left, ground truth, noisy measurements, and the two affine models making up the fitted mixture model are plotted.
    On the right, the probability $P(z_i=j)$ of each sensor reading (sorted by measurement location) to belong to the corresponding cluster $j\in\{1,2\}$ is depicted.}
    \label{fig:ill-res_2_clusters}
\end{figure}

Fig.~\ref{fig:ill-res_2_clusters} shows on the left the physical reality $f$ and the artificial measurement data as well as the fitted mixture model, and on the right the conditional probabilities $ P(z_i=j)$ from~\eqref{eq:z-posterior} of each sensor reading $i$ to belong to cluster $j$ given data and parameters after the final iteration of the EM algorithm. 
From that clustering we can infer that the affine model's parameters clearly depend on the sensor location $\xi$, with decreasing slope and increasing intercept of the locally fitted affine model. 
The high confidence in cluster assignment in the regions $\xi\in[0,0.3]\cup[0.7,1.0]$ could suggest a biphasic system behavior such as an ideal elasto-plastic stress-strain curve. However, the relatively large intermediate region $\xi\in[0.4,0.7]$ with almost constant non-confidence in assignment suggests a smoother transition. Unless at this point the modeler recognizes a so far neglected physical effect that would match this behavior, such as plasticity, one might formulate an improved phenomenological model with position-dependent slope simply by a quadratic dependency instead of an affine one: $\tilde f(\xi,a,b,c) = a\xi + b + c\xi^2$. This is not the true model, but no longer deficient.

\begin{table}
    \centering
    \begin{tabular}{cccc}
        \toprule
        cluster $j$  & weight $\omega_j$ &  slope $a_j$ & intercept $b_j$ \\
        \midrule
         1 & 0.41 & \num{0.40} & \num{0.67} \\
         2 & 0.59 & \num{1.66} & \num{0.03} \\
         \midrule
         1 & 0.00 & \num{0.00} & \num{0.00} \\
         2 & 0.59 & \num{1.66} & \num{0.03} \\
         3 & 0.41 & \num{0.40} & \num{0.67} \\
         4 & 0.00 & \num{0.00} & \num{0.00} \\         
         \bottomrule
    \end{tabular}
    \caption{Parameters identified by the EM algorithm for two and four clusters for the illustrative example.}
    \label{tab:parameters-ex1}
\end{table}

\begin{figure}
    \centering
    \includegraphics[width=\linewidth]{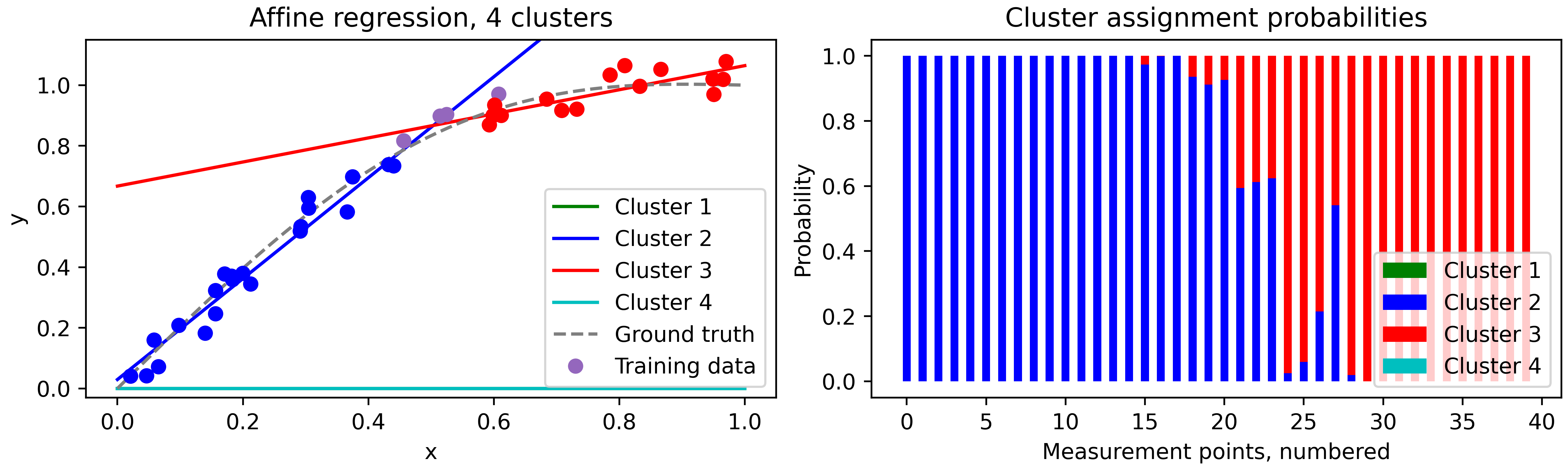}
    \caption{Results after the EM algorithm terminated for the illustrative example with 4 clusters. 
    The pictures depict as in Figure~\ref{fig:ill-res_2_clusters} the different models on the left, and the cluster assignment probabilities on the right.}
    \label{fig:ill-res_4_clusters}
\end{figure}

A more flexible mixture model with four clusters yields the same results. Tab.~\ref{tab:parameters-ex1} and Fig.~\ref{fig:ill-res_4_clusters} show that two clusters are essentially eliminated from the model, having a negligible weight $\omega_j\approx 0$.
The remaining two clusters with larger weight are the same as in the mixture model with only $k=2$ clusters. 
In this illustrative example, a larger number of clusters just increases model complexity and computing time. Of course, the marginal posterior~\eqref{eq:marginal-posterior} is highly nonconvex and usually has many local maximizers to which the EM algorithm can converge. It is conceivable that different starting parameters could have provided more balanced weights $\omega_j$ and a smoother transition of slopes $a_j$ and intercepts $b_j$,  which then would have suggested a continuous dependence on $\xi$ even stronger than the two-cluster mixture model. Such a result is more likely if the hyper-prior~\eqref{eq:hyper-prior} favors equi-sized clusters, i.e., for $\gamma \gg 1$.
On the other hand, overfitting to noisy data is more likely to occur, and not necessarily prevented by the hyper-prior. Thus, the example supports the above recommendation of startint the investigation with a small number $k$ of clusters, most often $k=2$.

\subsection{Heat transfer within a concrete bridge} \label{sec:bridge}

As a second example, we consider a more realistic physical system: the evolution of the temperature distribution within a concrete bridge. We present both detection of an artificial model deficiency caused by a deliberate model simplification, and detection of a real-world model deficiency from actual measurements.

\subsubsection{Thermal model} \label{sec:thermal-model}

\begin{figure}
    \centering
    \includegraphics[width=0.7\linewidth]{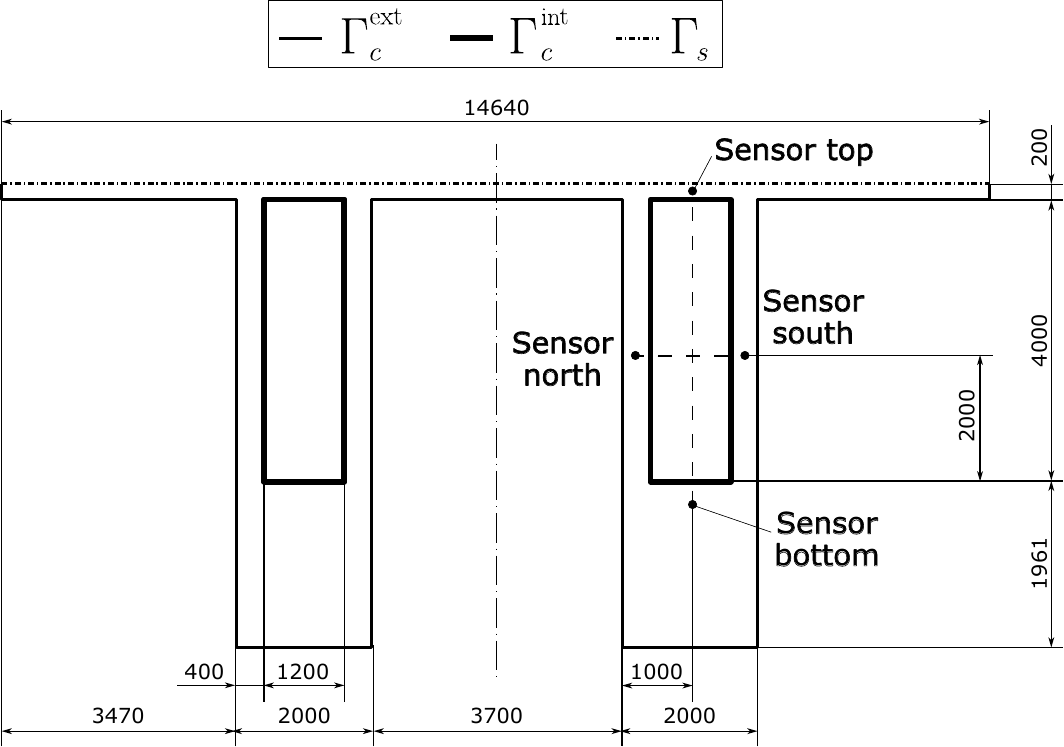}
    \caption{Cross-section of the 2D model used for the thermal bridge. Dimensions in mm. Measured perpendiculary from the inner hole surfaces $\Gamma_c^{\rm int}$ towards the concrete, sensor top is 5 mm deep, sensor north and sensor south are 50 mm deep and sensor bottom is 200 mm deep. }
    \label{fig:cross-section}
\end{figure}

Here, we still assume a known physical reality in the form of a 2D transient heat equation on a cross section of the bridge, numerically evaluated by a finite element simulation on a fixed grid until time $t_f$. Let $\Omega \subset \mathbb{R}^2$ be a bounded domain with boundary $\partial\Omega = \Gamma^{\mathrm{int}}_c\cup \Gamma^{\mathrm{ext}}_c \cup \Gamma_s$, where $\Gamma^{\mathrm{int}}_c$ is the boundary exposed to internal heat convection, $\Gamma^{\mathrm{ext}}_c$ is the boundary exposed to external heat convection and wind advection, and $\Gamma_s$ is the part exposed to convection, advection, and shortwave (solar) radiation, see Fig.~\ref{fig:cross-section}. 

We seek the temperature field $T: \Omega \times (0, t_f] \to \mathbb{R}$ such that
\begin{align} 
    \rho c_p \frac{\partial T}{\partial t} &= \mathrm{div}(\kappa \nabla T)  &&\text{in } \Omega \times (0, t_f], \label{eq:heat_eq_kappa}
\end{align}
where $\rho$ is the mass density [\si{\kilogram\per\metre\cubed}], $c_p$ is the specific heat capacity [\si{\joule\per\kilogram\per\kelvin}], $\kappa$ is the thermal conductivity [\si{\watt\per\metre\per\kelvin}] and $T$ is the temperature [\si{\kelvin}]. In a homogeneous material, it is convenient to define the thermal diffusivity $\alpha := \frac{\kappa}{\rho c_p}$ [\si{\metre\squared\per\second}]. Then,~\eqref{eq:heat_eq_kappa}  simplifies to
\begin{align}
    \frac{\partial T}{\partial t} &= \mathrm{div}(\alpha \nabla T) \quad &&\text{in } \Omega \times (0, t_f], \label{eq:heat_eq}
\end{align}
subject to the initial condition $T(x,0) = T_0 \text{ in } \Omega$ and the boundary conditions
\begin{align}
    -\alpha\rho c_p \nabla T \cdot \mathbf{n} &= c_c h_{\rm int} \left( T - T^{\mathrm{int}}_{\infty}(t) \right) &&\text{on } \Gamma^{\mathrm{int}}_c \times (0, t_f], \label{eq:convection_int} \\
    -\alpha\rho c_p \nabla T \cdot \mathbf{n} &= c_c h_{\rm ext} \left( T - T^{\mathrm{ext}}_{\infty}(t) \right) &&\text{on } \Gamma^{\mathrm{ext}}_c \times (0, t_f], \label{eq:convection_ext} \\
    -\alpha\rho c_p \nabla T \cdot \mathbf{n} &= c_c h_{\rm ext} (T - T^{\mathrm{ext}}_\infty(t)) - c_r a I_{\text{in}}(t) \quad &&\text{on } \Gamma_s \times (0, t_f], \label{eq:shortwave}
\end{align} 
where $\mathbf{n}$ is the outward unit normal to $\partial\Omega$, $h_{\rm ext}$ is the convective heat transfer coefficient [\si{\watt\per\metre\squared\per\kelvin}] at the exterior of the bridge, $h_{\rm int}$ is the convective heat transfer coefficient [\si{\watt\per\metre\squared\per\kelvin}] at the interior of the bridge, $T_\infty(t)$ is the time-dependent ambient temperature [\si{\kelvin}], $a$ is the surface absorptivity (dimensionless), $I_{\text{in}}(t)$ is the time-dependent incident shortwave irradiance [\si{\watt\per\metre\squared}], $T_0$ is the homogeneous initial temperature distribution, $c_c$ is a dimensionless joint correction factor for all convection boundaries, and $c_r$ is a dimensionless correction factor for shortwave radiation. 

The model parameters considered for identification are the diffusivity $\alpha$ and the multiplicative convection coefficient correction $c_c$ for the natural convection boundary term in the experiment described in Sec.~\ref{exp:gen-data}, while in the experiment from Sec.~\ref{exp:bridge-monitoring} we also infer the shortwave irradiation coefficient $c_r$.
The considered domains and priors can be found in Tab.~\ref{tab:inference-parameters}; a priori, the parameters are assumed to be independent from one another.

\begin{table}
    \centering
    \begin{tabular}{ccc}
    \toprule
    Parameter        & Domain & Prior\\
    \midrule
    $\alpha$         & $[0.5,5]\,\si{\milli\metre\squared\per\second}$          & $\mathcal{N}(1,0.2)$\si{\milli\metre\squared\per\second} \\
    $c_c$            & $[0.1 ,10^2]$                                            & Uniform$([0.1 ,10^2])$ \\
    $c_r$           & $[0.01 ,10]$                                              & Uniform$([0.01 ,10])$ \\
    \bottomrule
    \end{tabular}
    \caption{Domains and priors for the inference parameters in the thermal bridge model.}
    \label{tab:inference-parameters}
\end{table}

\begin{table}
    \centering
    \begin{tabular}{crl}
    \toprule
    Parameter        & Value & \\
    \midrule
    $h_{\rm int}$    & 10.00   &\si{\watt\per\metre\squared\per\kelvin} \\
    $h_{\rm ext}$    & 14.10  &\si{\watt\per\metre\squared\per\kelvin} \\
    $\rho$           & 2400 &\si{\kilogram\per\metre\cubed} \\
    $c_p$            & 870  &\si{\joule\per\kilogram\per\kelvin} \\
    $a$              & 0.275& \\
    \bottomrule
    \end{tabular}
    \caption{Values of the parameters with assigned value in the thermal bridge model.}
    \label{tab:fixed-parameters}
\end{table}

The values of the parameters not considered for inference are given in Tab.~\ref{tab:fixed-parameters}.
To calculate the external heat transfer coefficient $h_{\text{ext}}$, we start by computing the Reynolds number using the wind speed, characteristic length, and air viscosity as $\mathrm{Re} = \frac{U \cdot L}{\nu}$ where $U$ is the wind speed, $L = \SI{4.0}{\metre}$ is the approximate characteristic length, and $\nu = \SI{1.81e-5}{\pascal\second}$ is the dynamic viscosity of air. The characteristic length for complex shapes can only be obtained experimentaly, therefore in this case the approximation is chosen arbitrarily between the value of $\SI{2.51}{\metre}$ obtained from the hydraulic diameter and $\SI{7.43}{\metre}$ of the chord lenght until half the cross-section of the bridge supposing an airfoil-like structure. The Nusselt number is then obtained from the empirical correlation $\mathrm{Nu} = 0.037 \, \mathrm{Re}^{4/5} \, \mathrm{Pr}^{1/3}$ where the Prandtl number of air is $\mathrm{Pr} = 0.71$. Finally, the convective heat transfer coefficient is calculated as $h_{\text{ext}} = \frac{\mathrm{Nu} \cdot k}{L} \approx \SI{14.10}{\watt\per\metre\squared\per\kelvin}$ with $k = \SI{0.025}{\watt\per\metre\per\kelvin}$ representing the thermal conductivity of air.

For the numerical simulation of the temperature evolution\added{,} we employ continuous second order finite elements on an unstructured but approximately uniform mesh consisting of 2611 vertices and 4462 triangles for space discretization, and an implicit Euler scheme with a step size of \SI{4}{\hour} for time discretization, consistent with the input sampling rate.

\subsubsection{Monitoring data}\label{sec:real-world-data}
The model of Sec.~\ref{sec:thermal-model} represents the Nibelungenbrücke in Worms (Germany). The bridge is equipped with a monitoring system~\cite{eisermannNBDescription2024, kangNBMonitoringDescription2024} providing measurements of the environmental conditions. It comprises, among other components, sensors measuring the external air temperature at the bridge site, denoted by $T_\infty^{\mathrm{ext}}$, the internal air temperature within the bridge structure, $T_\infty^{\mathrm{int}}$, the shortwave solar irradiation incident on the bridge deck, $I_{\rm in}$, and the relative humidity at the bridge location, $\mathrm{RH}$. These variables are recorded at a sampling frequency of 10~Hz. For the purpose of this study, the data are subsampled to one measurement every four hours via averaginga, and subsequently employed in the computation of thermal loads.

Measurements of the structural temperature are available from sensors positioned as shown in Fig.~\ref{fig:cross-section}. These are utilized in Sec.~\ref{exp:bridge-monitoring}. As illustrated, the sensors are distributed around the inner wall of one of the box girders in proximity to the pilot. They are designated as bottom (b), top~(t), south (s), and north (n) sensor, respectively. The distances of the sensors from $\Gamma_c^{\rm int}$ have been corrected by model calibration from their original position specification. We use distances of 20 cm for the bottom sensor, 1.5 cm for the top sensor, 5 cm for the south sensor, and 15cm for the north sensor.

\subsubsection{Missing irradiation as artificial model deficiency}\label{exp:gen-data}
In this example, we test the ability of mixture models to point at model deficiency causes by deliberately simplifying the thermal bridge model from Sec.~\ref{sec:thermal-model}. We generate simulated measurement data from the full model with a solar irradiation source term ($c_r=1.0 \cdot10^{-1}$) and fit a model which does not account for shortwave irradiation ($c_r=0$). 

Temperature sensor measurement data are simulated using air temperatures $T_\infty^{\rm int} = T_\infty^{\rm ext}$ from June 2024, diffusivity $\alpha = \SI{1.0}{\milli\metre\squared\per\second}$, a convection correction factor $c_c = 5.0$, and an initial homogeneous temperature $T_0 = \SI{20}{\celsius}$ for the structure. The previous 180 steps, i.e., one month, are evaluated and discarded, allowing the structure to develop a heterogeneous temperature profile. The measurements consist of the bridge's simulated temperature in the four sensor locations measured every four hours, for a total of $n = 30\cdot 4 \cdot 6= 720$ sensor readings, perturbed with independent normally distributed additive noise with mean zero and standard deviation $\sigma = \SI{0.05}{\kelvin}$.

The base model using the maximum posterior point estimate of a single parameter vector is obviously deficient, with a KS statistic $D=0.99$ and deviating parameter values, as given in Tab.~\ref{tab:parameters-ex2}.
\begin{table}
    \centering
    \begin{tabular}{lcccS[table-format=3.2]S[table-format=4.2]cc}
        \toprule
        model  &  sensors & weight & diffusivity $\alpha$ & {convection $c_c$} & {deviation $\mathcal D$} &  KS statistic $D$ \\
        \midrule
         truth & all & & \SI{1.00}{\milli\metre\squared\per\second} & 5.00 & 0 & 0.00 \\
         \midrule
         base model & all &  & \SI{0.65}{\milli\metre\squared\per\second} & 18.37 & 1026.43 & \num{0.99} \\
         \midrule 
         cluster 1 & t     & 0.24 & \SI{1.71}{\milli\metre\squared\per\second} & 100.00 & 203.98 & \num{0.97} \\
         cluster 2 & b,n,s & 0.76 & \SI{1.12}{\milli\metre\squared\per\second} & 5.19 & 1.00 & \num{0.23} \\
         mixture model & all & & & & 2.42 & \num{0.35} \\
         \bottomrule
    \end{tabular}
    \caption{Results of maximum posterior parameter estimates for the thermal bridge model ignoring irradiation. For the individual clusters of the mixture model, the discrepancy is computed on the readings assigned to the respective cluster.}
    \label{tab:parameters-ex2}
\end{table}
For a mixed model, we consider $k=2$ clusters and, as the computational costs of running the simulations is not negligible and empirically the clusters' parameters stabilize after around 8 iterations, the EM algorithm performs 10 iterations only.
For the M-step, we use the setup described at the beginning of Sec.~\ref{sec:exp}.
\begin{figure}
    \centering
    \includegraphics[width=\linewidth]{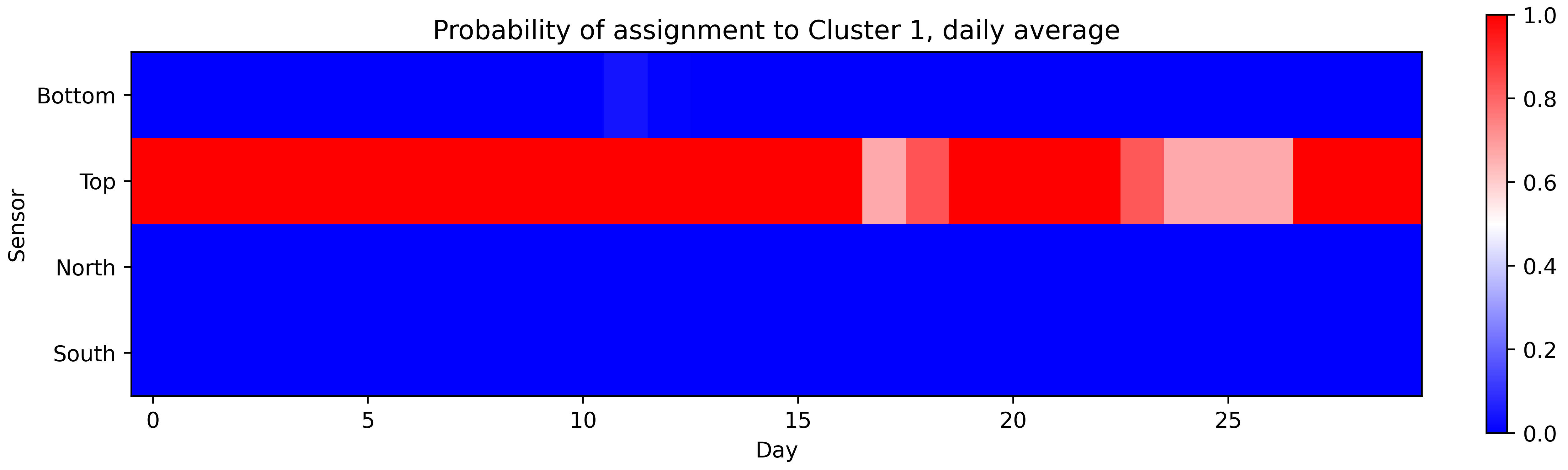}
    \caption{Average over each day of the probability $\bar p$ for a sensor reading to belong to a cluster for the thermal bridge with generated data example. }
    \label{fig:simul-probabilities}
\end{figure}
Fig.~\ref{fig:simul-probabilities} depicts the assignment of sensor readings to clusters: the mixture model separates the top sensor from the three others, very clearly for almost all time points. This is consistent with the known model deficiency, as the solar irradiation acts on the top boundary $\Gamma_s$ only.

\begin{figure}
    \centering
    \includegraphics[width=\linewidth]{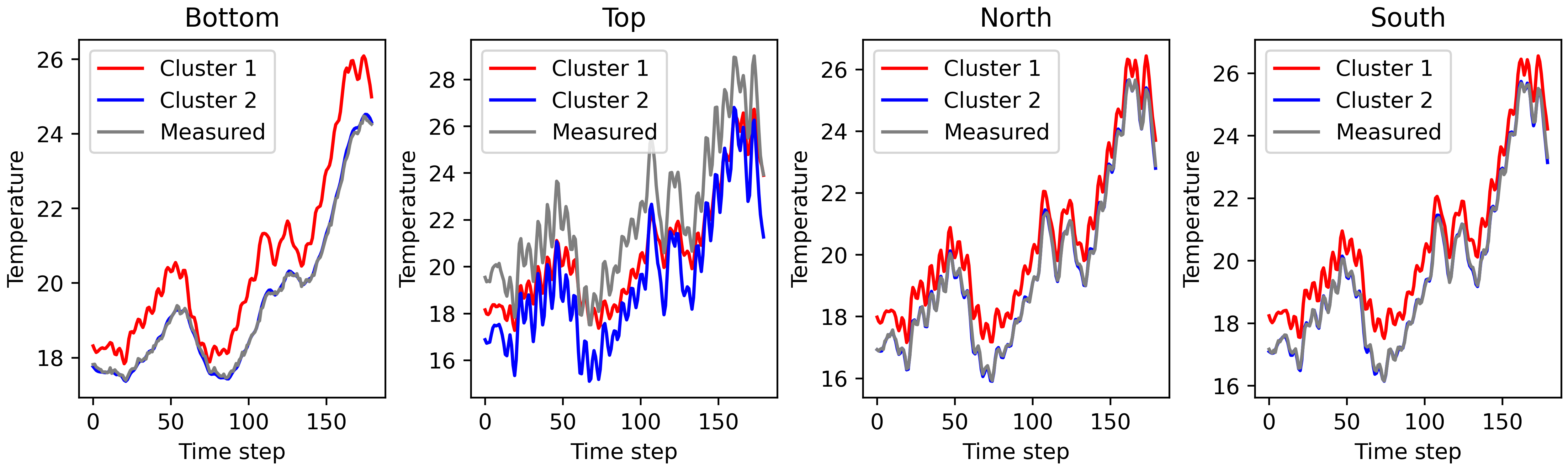}
    \caption{Temperature time series for the two clusters and measurement in each of the 4 sensors  for the thermal bridge with generated data example.}
    \label{fig:simul-temperature}
\end{figure}

The parameter values identified for each cluster are given in Tab.~\ref{tab:parameters-ex2}.
For cluster 2, they are  reasonably close to the ground truth, and 
the model deviation, when considering only those readings assigned to cluster 2, is quite small. In contrast, in cluster 1 the convection coefficient assumes the extreme value $c_c = \num{1e2}$ within the admissible range and the model deviation is large. This suggests that even with extreme parameter values, the simplified model is unable to reproduce the top sensor readings. 
This can also be seen in the temperature time series plotted in Fig.~\ref{fig:simul-temperature}, where in the top sensor the model output is considerably off the measurement data for both clusters, while cluster 2 matches the sensor readings adequately in the other sensors. The whole mixture model performs quite well, achieving a deviation in between the two clusters and much smaller than the base model. The mixture model could therefore serve as an improved model, but since it does not provide any spatial distribution of diffusivity and convection, it also does not allow to predict temperature values at a new sensor location. We thus turn to physically interpretable model improvement.

Based on the discussion of possible model deficiency causes and their impact on identified mixture model parameters in Sec.~\ref{sec:interpretation}, we might assume that some neglected effect drives the system into a region of the state space that is not reachable by the model. Given that the temperatures produced by the parameter vector of cluster 1 are generally above the ones predicted by the parameter vector of cluster 2, but still do not reach the level of top sensor readings, we may conclude that there is some heat source acting close to the top sensor. Solar irradiation is then an obvious candidate.

In fact, the neglected irradiation introduces thermal energy into the structure, increasing its temperature in the top part above the temperatures that can be reached by heat conduction. The latter is restricted to the range of the exterior temperatures, and leads to time-averaged sensor readings. No modification of diffusivity $\alpha$ or heat exchange $c_c$ with the environment can raise the temperature level over a longer period of time. The parameter optimization for cluster 1 nevertheless tries to achieve this by increasing the convection factor $c_c$ and diffusivity $\alpha$, allowing the slowly increasing exterior temperature to diffuse faster into the bridge body, and in particular to the top sensor location close to the boundary. 

\subsubsection{Full model discrepancy from monitoring data}\label{exp:bridge-monitoring}

In the third example, we aim at identifying diffusivity $\alpha$, convective heat transfer correction $c_c$, and the shortwave irradiation coefficient $c_r$ for the full model of Sec.~\ref{sec:thermal-model} from actual sensor readings provided by the monitoring system of the Nibelungen bridge.

The measurements on the structure are taken with the measurement setup described in Sec.~\ref{sec:real-world-data}, consisting of four sensors measuring the bridge's temperature every four hours for the 30 days of June 2024, assuming additive normal noise $\mathcal N (0, \sigma^2)$ with $\sigma = \SI{0.05}{\kelvin}$. The priors and domains for the parameters to be identified are given in Table~\ref{tab:inference-parameters}.

We start again with $k=2$ clusters in the mixture model.
The Finite Element simulation is computationally expensive as in the preceding example: the number of iterations for the EM algorithm is again limited to 10, and the M-step is performed as introduced at the beginning of Sec.~\ref{sec:exp}. The identified cluster mean parameters are provided in Tab.~\ref{tab:parameters-ex3}. In contrast to the previous example of omitted irradiation, both clusters show reasonable parameter values within the admissible range.

\begin{table}
    \centering
    \begin{tabular}{ccccc}
        \toprule
        cluster $j$ &   weight $\omega$ & diffusivity $\alpha$ & convection $c_c$ & irradiation $c_r$\\
        \midrule
         1 & 0.36 & \SI{1.65}{\milli\metre\squared\per\second} & \num{26.16} & \num{0.36}  \\
         2 & 0.64 & \SI{0.92}{\milli\metre\squared\per\second} & \num{6.20} & \num{0.11}  \\
         \bottomrule
    \end{tabular}
    \caption{Identified parameters for two clusters of fitting the full thermal bridge model to real-world data. }
    \label{tab:parameters-ex3}
\end{table}

\begin{table}
    \centering
    \begin{tabular}{lcc}
        \toprule
        Model  & Deviation $\mathcal D$ &  KS statistic $D$ \\
        \midrule
         Base model & 175.29 & \num{0.91} \\
         Mixture model & 18.60 & \num{0.68}  \\   
         Humidity model & 30.45  & \num{0.75}\\
         \bottomrule
    \end{tabular}
    \caption{Model discrepancy for the thermal bridge with real world data example. 
    The parameter $\param_*$ where the discrepancy is evaluated is the Maximum a Posteriori estimate. 
    }
    \label{tab:discrepancy-ex3}
\end{table}

\begin{figure}
    \centering
    \includegraphics[width=\linewidth]{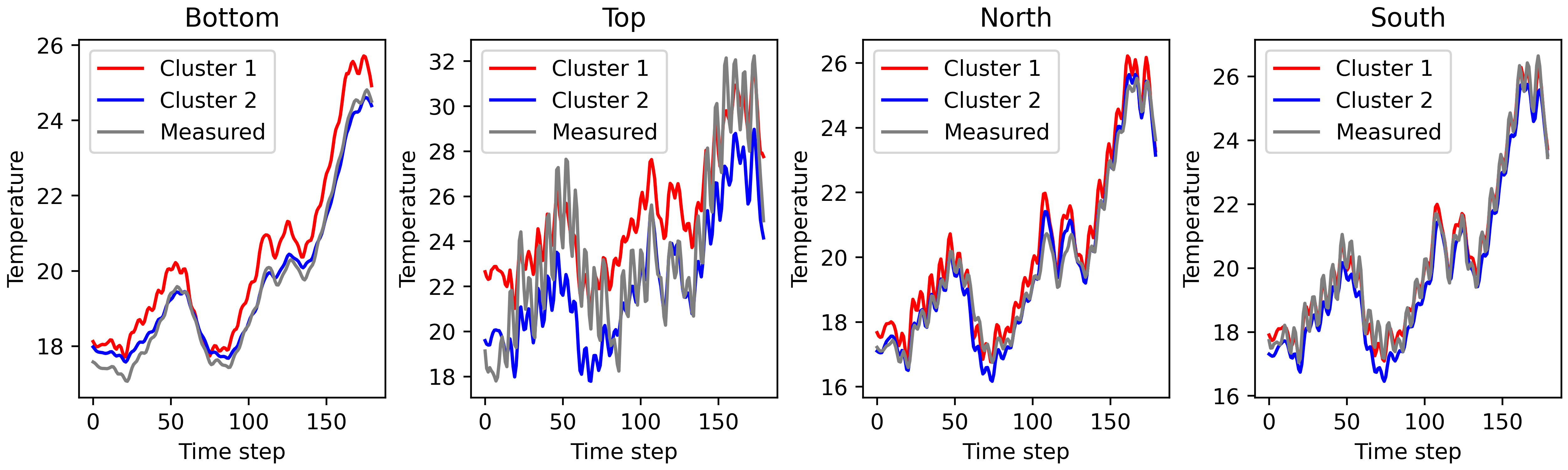}
    \caption{Temperature time series for the two clusters and measurement in each of the 4 sensors for the thermal bridge with real-world data example.}
    \label{fig:real-temperature}
\end{figure}

\begin{figure}
    \centering
    \includegraphics[width=\linewidth]{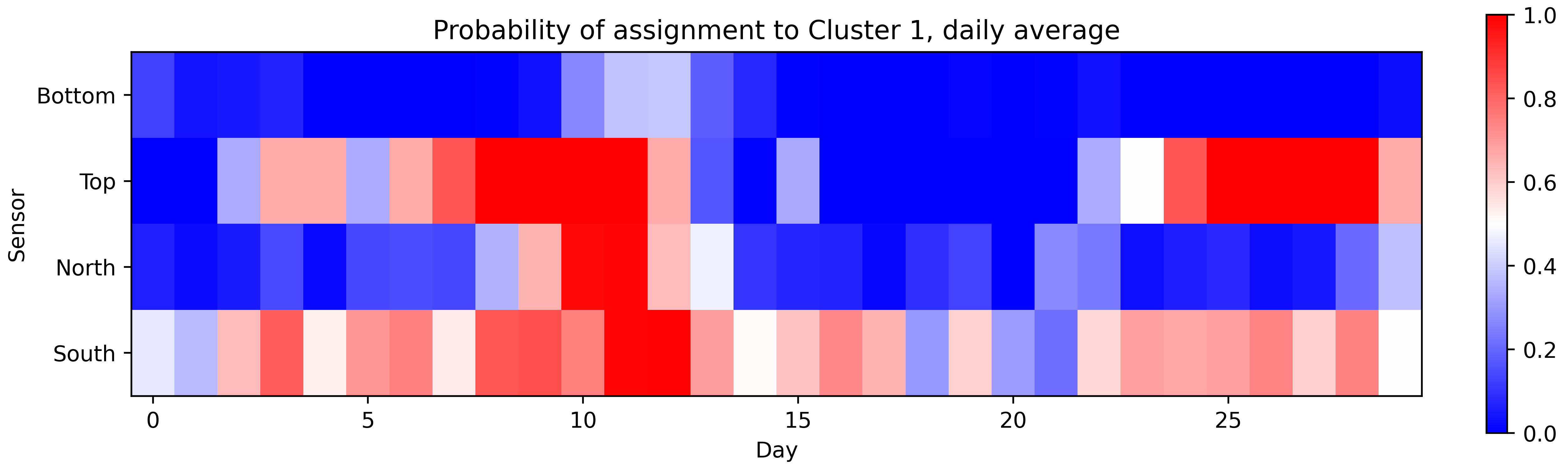}
    \caption{Average over each day of the probability $\bar p$ for a sensor reading to belong to a cluster for the thermal bridge with real-world data.}
    \label{fig:real-probabilities}
\end{figure}

Fig.~\ref{fig:real-temperature} plots the temperature time series for the two clusters and the measurements for this example.
We can notice that in the top sensor, the temperature oscillates significantly more than in other sensors and the two clusters match different parts of the sensor b time series.
This can also be noticed in the average daily cluster assignment probability in Fig.~\ref{fig:real-probabilities}, where the top sensor displays the clearest temporal structure in assignment to one cluster or the other. 

\begin{figure}
    \centering
    \includegraphics[width=0.5\linewidth]{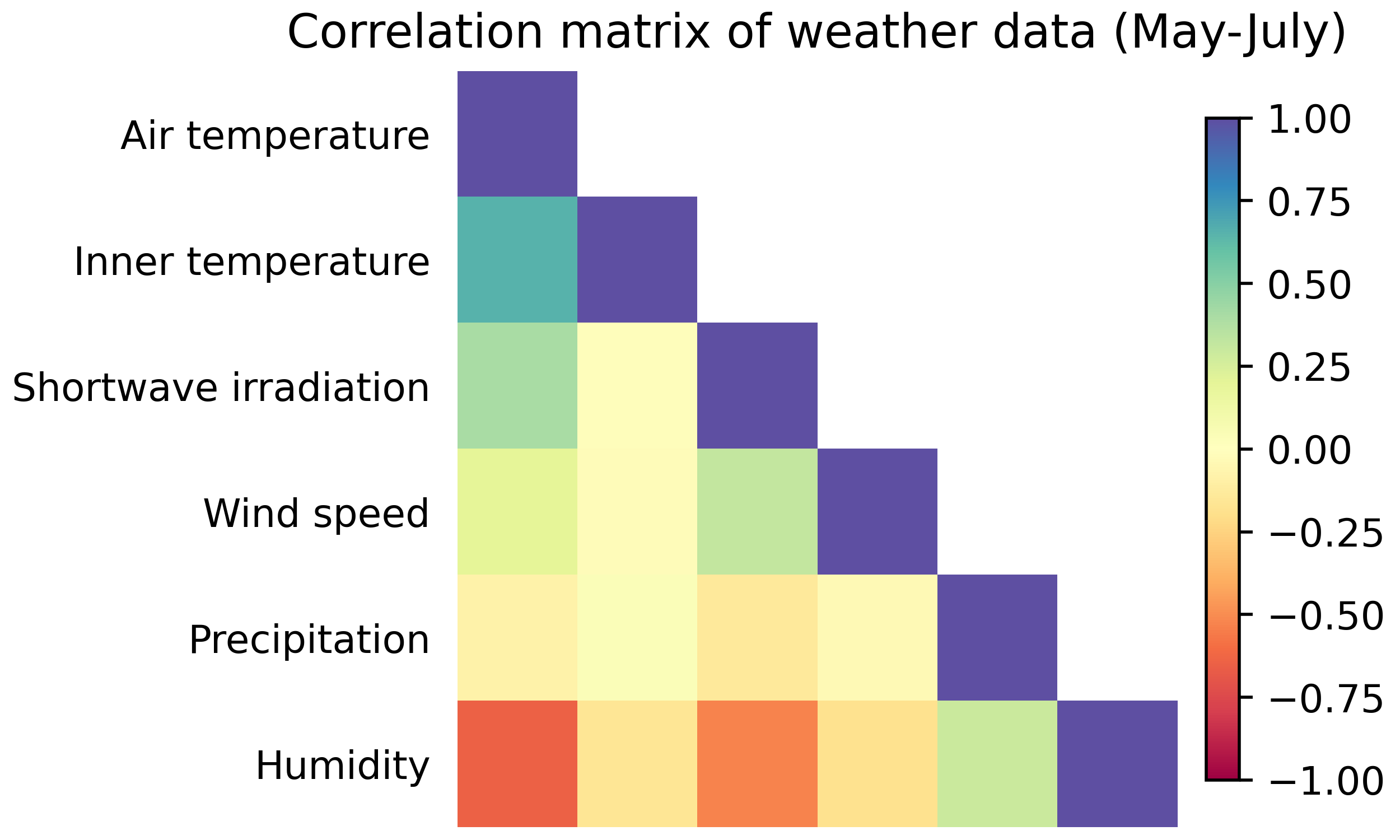}
    \caption{Correlation between the considered environmental factors, average every 4 hours of the data over the three months of May, June and July 3}
    \label{fig:weather-correlations}
\end{figure}

In this example, the deficiencies of the model are unknown. The pronounced time dependence of the cluster assignment in particular for the top sensor, but, to some extent, also for the other sensors, suggests a time-dependent variation of parameters. As the structure itself is unlikely to change repeatedly on a time scale of days, variable environmental conditions are a strong candidate. The exterior temperature $T_\infty^{\rm ext}$ and irradiation $I_{\rm in}$ time courses have already been taken into account. We therefore consider further available weather data over three months: humidity, precipitation, interior temperature $T_\infty^{\rm int}$, wind speed and direction, and cloud covering. The correlation matrix of available weather variables is plotted in Fig.~\ref{fig:weather-correlations}.

\begin{figure}
    \centering
    \includegraphics[width=\linewidth]{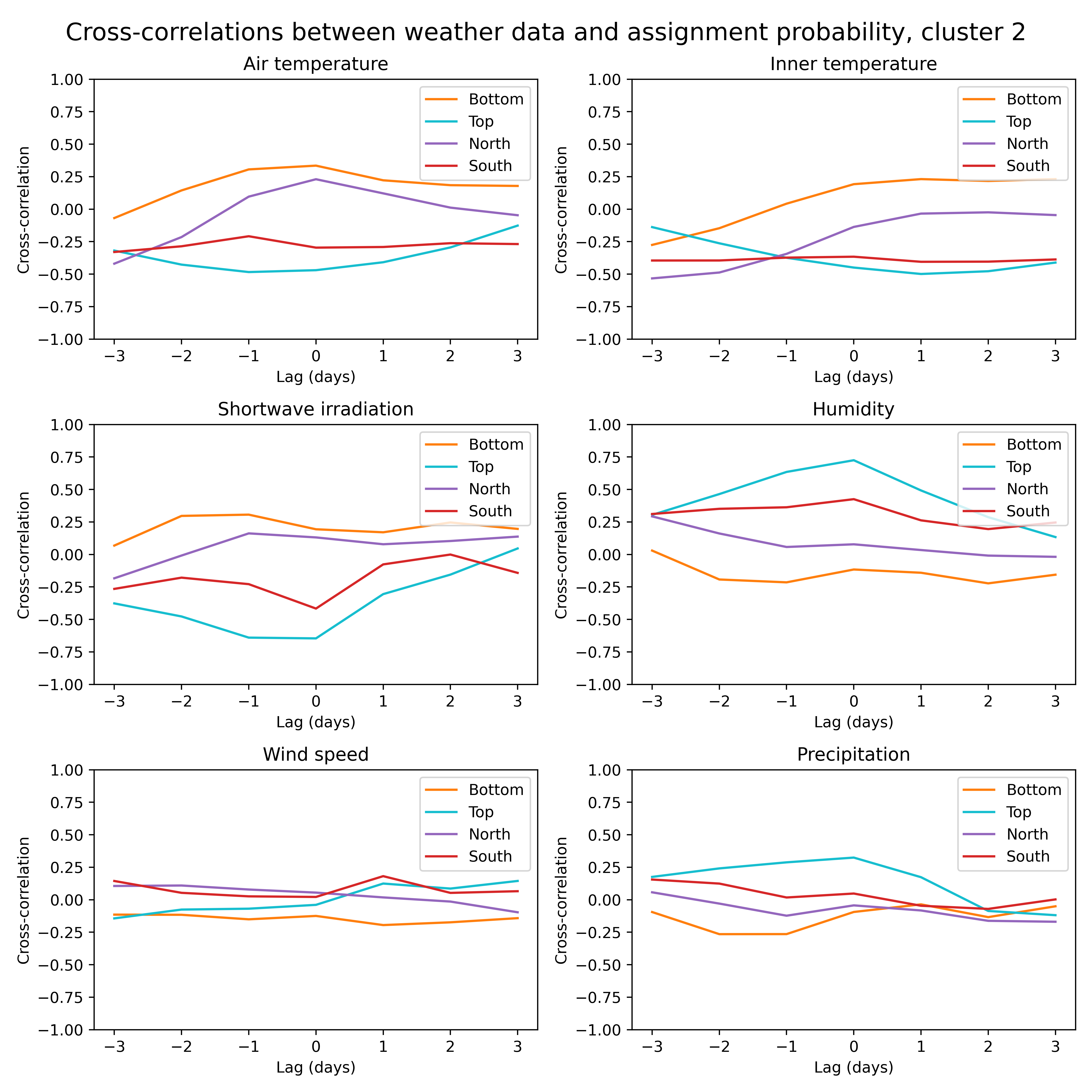}
    \caption{Cross correlations between the time series of the assignment probabilities of the different clusters and the time series of different environmental factors, daily average of the data over the three months of May, June and July 2024. }
    \label{fig:cross-correlations}
\end{figure}

\begin{figure}
    \centering
    \includegraphics[width=\linewidth]{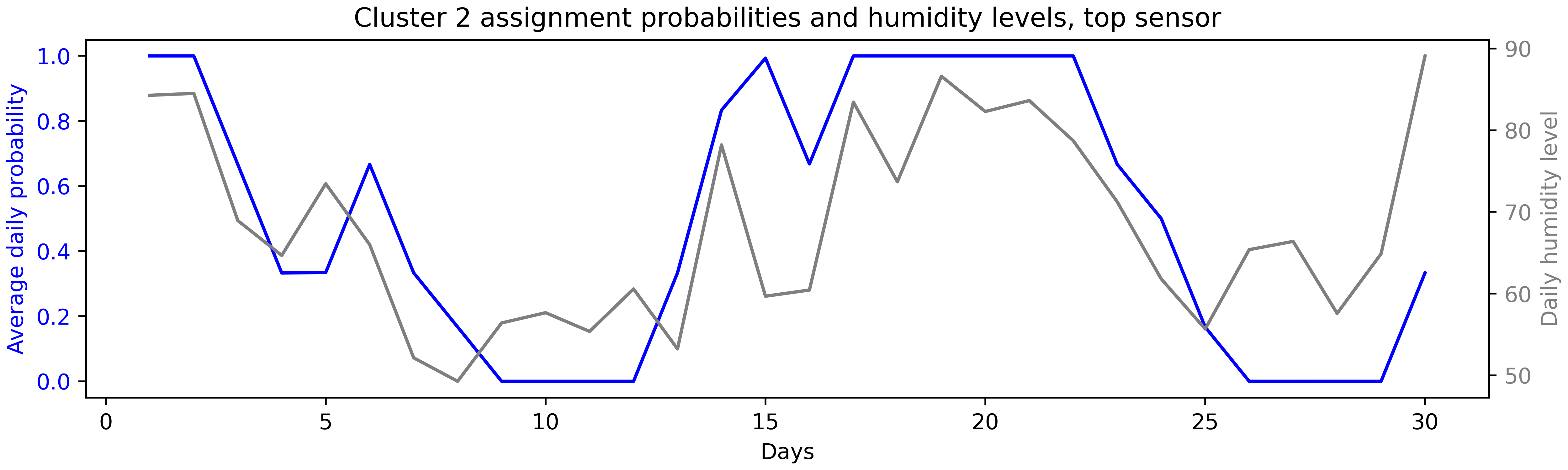}
    \caption{Average daily cluster probability $\bar p_{\cdot, 2} $ of belonging to cluster 2 for the readings from the top sensor in the real-world data example, along with average daily humidity level for the month of June 2024.}
    \label{fig:real-humidity}
\end{figure}

We compute the correlation of these weather data with the cluster assignment of the four sensors for different time shifts between weather and cluster assignment. For that, we use daily averaged data to avoid strong circadian correlations dominating the results. The strongest correlations are shown in Fig.~\ref{fig:cross-correlations}. 
We observe that humidity correlates most with the cluster assignment, in particular for the top sensor, which also shows the most pronounced switching between clusters. The time series of humidity and cluster 2 assignment probabilities over the month of June 2024 are depicted in Fig.~\ref{fig:real-humidity}. The correlation is strongest for a zero time delay, but still large for $\pm \SI{1}{\day}$. Results of a finer time delay analysis for humidity, reported in Tab.~\ref{tab:humidity-corrs}, reveal a delay of \SI{4}{\hour} to yield the largest correlation between humidity and sensors top and south. As high humidity corresponds to lower temperatures, we conjecture that evaporation leads to a loss of heat or that surface wetting decreases the emissivity and therefore the irradiation factor $c_r$.

We therefore extend the model's boundary condition~\eqref{eq:shortwave} with an additional term that depends on the relative humidity $\mathrm{RH}$:
\begin{equation}
    -\alpha\rho c_p \nabla T \cdot \mathbf{n} = c_c h_{\rm ext} (T - T^{\mathrm{ext}}_\infty(t)) - c_r a I_{\text{in}}(t)-c_\mathrm{RH}\mathrm{RH}(t) \quad \text{on } \Gamma_s \times (0, t_f].
\end{equation}
Here, $c_\mathrm{RH}$ is a dimensionless correction factor set to $10^{-4}$ and not further calibrated in the present work. This results in an improvement of the model, as shown in Tab.~\ref{tab:discrepancy-ex3}: while the model with the humidity factor does not perform as well as the mixture model with two sets of parameters, it reduces the model deviation $\mathcal D$ by a factor of six when compared to the base model. The model deficiency remains high, however, suggesting that further improvement is needed, but a first, apparently dominant, and physically interpretable factor of model discrepancy has been identified.

\begin{table}
    \centering
    \begin{tabular}{lccccc}
    \toprule
        Sensor &  \multicolumn{5}{c}{Lag w.r.t. humidity} \\
          & -8h  & -4h & 0h & 4h & 8h\\
         \midrule
         Top & \textbf{0.65} & \textbf{0.65} & 0.37 & 0.10 & 0.05 \\
         Bottom &  -0.20 & -0.16 & -0.05 & 0.02 & 0.00\\
         North & -0.01 & -0.14 & -0.12 & 0.06 & 0.21 \\
         South & 0.40 & \textbf{0.65} & 0.41 & -0.04 & -0.29\\
         \bottomrule
    \end{tabular}
    \caption{Cross correlations between the time series of humidity and the time series of assignment probabilities for cluster 2, average every 4 hours of the data over the three months of May, June and July 2024. Small deviations from Fig.~\ref{fig:cross-correlations} are due to the finer resolution of lag times considered here.}
    \label{tab:humidity-corrs}
\end{table}

\section{Conclusion and outlook}
In this work, we develop a strategy for detecting model deficiency and its possible causes that is based on mixture models. It clusters available sensor readings to different parameter vectors, with the aim of recovering a meaningful structure within the clustering and identified parameter values. This can help the modeler in identifying causes of model deficiencies and structural biases, and ultimately in improving the model by inclusion of so far neglected physical processes that appear to be related to the observed mixture model structure. The approach is validated on examples of increasing complexity: in the final one, we identify an unknown effect on a real-world model using real-world data. 
While some reasons for model deficiency might not be easily detected by the proposed strategy, such as spatial inhomogeneities of parameters that affect many sensor locations in a similar way, the numerical experiments suggest that the approach can be effective in detecting localized deficiencies, time-related phenomena, and nonlinearity.

The method is flexible in the choice of number of clusters as well as in the related choice of the hyper-prior parameter $\gamma$. While adjusting those parameters can provide additional information on the structure of a model deficiency, there is no natural way for selecting them. The proposed heuristic of starting simple, with few clusters, is plausible and works well in the presented examples, but requires more investigation. 

One limitation is the need for interpreting the clustering structure and identified parameter vectors' values. While we provide heuristic arguments for expected relations between mixture model structure and possible causes of model deficiency that can guide model improvement, the actual formulation of an improved model is still left to the modeler, is application-specific, and requires a substantial amount of modeling expertise. The reason is, that the mixture model approach provides a very coarse approximation of, e.g., parameter dependencies or nonlinearities. One promising direction for future research is therefore a more explicit representation and identification of such parameter dependencies.

\section*{Acknowledgement}
This work has been funded by the German Research Foundation (DFG) under grant 501811638 within the project C07 of the priority program SPP 2388 "Hundert plus – Verlängerung der Lebensdauer komplexer Baustrukturen durch intelligente Digitalisierung" in Phase I and under grant 562812195 within the project A04 of the same program in Phase II.
We thank the collaboration of Marx Krontal Partner (MKP GmbH.) and Landesbetrieb Mobilität Worms (LBM Worms) for supplying the data of the Nibelungenbrücke.

\bibliographystyle{abbrv}
\bibliography{references}

\end{document}